\documentclass[12pt]{iopart}

\newcommand{\vr}{{\bf{r}}}
\newcommand{\vp}{{\bf{p}}}
\newcommand{\K}{{\mathcal{K}}}
\newcommand{\A}{{\mathcal{A}}}
\renewcommand{\P}{{\mathcal{P}}}
\newcommand{\lL}{{l_\mathrm{L}}}
\newcommand{\lD}{{l_\mathrm{d}}}
\newcommand{\lF}{{l_\mathrm{f}}}
\newcommand{\su}{{|\!\uparrow\rangle}}
\newcommand{\sd}{{|\!\downarrow\rangle}}

\usepackage{epsfig}
\usepackage{amssymb}
\usepackage{setstack}
\usepackage{revsymb}

\begin{document}

\title[Loschmidt echo for local perturbations]{Loschmidt echo for local perturbations:\\
  non-monotonous cross-over from the Fermi-golden-rule to the escape-rate
  regime}

\author{Arseni Goussev$^{1,2}$, Daniel Waltner$^1$, Klaus Richter$^1$ and Rodolfo A Jalabert$^{1,3}$}

\address{$^1$ Institut f\"ur Theoretische Physik, Universit\"at Regensburg, 93040 Regensburg, Germany}
\address{$^2$ School of Mathematics, University of Bristol, University Walk, Bristol BS8 1TW, United Kingdom\footnote{Present address}}
\address{$^3$ Institut de Physique et Chimie des Mat{\'e}riaux de Strasbourg, UMR
  7504 (CNRS-ULP), 23 rue du Loess, BP 43, 67034 Strasbourg Cedex 2, France\footnote{Present and permanent address}}

\ead{arseni.goussev@bristol.ac.uk}

\date{\today}

\begin{abstract}
  We address the sensitivity of quantum mechanical time evolution by
  considering the time decay of the Loschmidt echo (LE) (or fidelity) for
  local perturbations of the Hamiltonian. Within a semiclassical approach we
  derive analytical expressions for the LE decay for chaotic systems for the
  whole range from weak to strong local perturbations and identify different
  decay regimes which complement those known for the case of global
  perturbations. For weak perturbations a Fermi-golden-rule (FGR) type
  behavior is recovered. For strong perturbations the escape-rate regime is
  reached, where the LE decays exponentially with a rate independent of the
  perturbation strength. The transition between the FGR regime and the
  escape-rate regime is non-monotonic, i.e. the rate of the exponential
  time-decay of the LE oscillates as a function of the perturbation strength.
  We further perform extensive quantum mechanical calculations of the LE based
  on numerical wave packet evolution which strongly support our semiclassical
  theory. Finally, we discuss in some detail possible experimental realizations
  for observing the predicted behavior of the LE.
\end{abstract}

%\submitto{\NJP}

\maketitle

\section{Introduction}

One of the most prominent manifestations of chaos in classical
physics is the hypersensitivity of the dynamics to perturbations in the 
initial conditions or Hamiltonian. That is, two trajectories of 
a chaotic system launched from two infinitesimally close phase-space 
points deviate exponentially from each other; so do
the trajectories starting from the same point in phase space, but
evolving under slightly different Hamiltonians. In a quantum system it
is natural to consider $|\langle\phi_1|\phi_2\rangle|^2$
as a measure of ``separation'' of two quantum states $|\phi_1\rangle$
and $|\phi_2\rangle$. The unitarity of quantum propagators renders the
overlap of any two states of the same system unchanged in the course
of time.  Thus, quantum systems are said to be stable with respect to
perturbations of the initial state. However, a perturbation of the
Hamiltonian can (and usually does) result in a nontrivial
time dependence of the wave function overlap, suggesting a viable
approach for describing instabilities and, therefore, for quantifying
chaos in quantum systems.

Peres \cite{peres} proposed to consider the overlap
\begin{equation}
  O(t) = \langle\phi_0| e^{i\tilde{H}t/\hbar} e^{-iHt/\hbar}
  |\phi_0\rangle 
  \label{I1}
\end{equation}
of the state $e^{-iHt/\hbar} |\phi_0\rangle$, resulting from an initial
state $|\phi_0\rangle$ after evolution for a time $t$ under the
Hamiltonian $H$, with the state $e^{-i\tilde{H}t/\hbar} |\phi_0\rangle$ obtained
from evolving the same initial state through $t$, but under a slightly
different (perturbed) Hamiltonian $\tilde{H}$. He showed that the
long-time behavior of 
\begin{equation}
  M(t) = \left| O(t) \right|^2
  \label{LE}
\end{equation}
depends on whether the underlying classical dynamics is regular or chaotic.

In the field of quantum computing $M(t)$ is an important concept, usually
referred to as {\it fidelity} \cite{Chuang}. Moreover, $M(t)$ can be also
interpreted as the squared overlap of the initial state $|\phi_0\rangle$ and
the state obtained by first propagating $|\phi_0\rangle$ through time $t$
under the Hamiltonian $H$, and then through time $-t$ under the perturbed
Hamiltonian $\tilde{H}$ (or $-\tilde{H}$ from $t$ to $2t$).  This
time-reversal interpretation constitutes a description of the echo experiments
that have been performed by nuclear magnetic resonance since the fifties
\cite{cit-HannEcho}. When the Hamiltonian $H$ describes some complex
(many-body or chaotic) dynamics $M(t)$ is referred to as {\it Loschmidt echo}
(LE) \cite{cit-MolPhysics}, and this is the terminology we will adopt.

By construction, the LE equals unity at $t=0$, and typically decays further in
time. Most of the analytical studies so far addressed the quantity
$\overline{M(t)}$ corresponding to the LE averaged either over an ensemble of
initial states, or over an ensemble of different perturbed (and/or
unperturbed) Hamiltonians. $\overline{M(t)}$ has been predicted to follow
different decay regimes in various chaotic systems with several Hamiltonian
perturbations \cite{jalab,gorin}.  Depending on the nature and strength of the
perturbation, $\tilde{H}-H$, one recognizes the {\it perturbative} Gaussian
\cite{jacq-1,toms,pros}, the non-diagonal or {\it Fermi-golden-rule} (FGR)
\cite{jalab,jacq-1,toms,pros-NEW} and the diagonal or {\it Lyapunov}
\cite{jalab,cucch-1} regimes.  Here, `diagonal' and `non-diagonal' refer to
the underlying pairing of interfering paths in a semiclassical approach, see
Sec.~\ref{section_theory}. The perturbative, FGR and Lyapunov regimes, listed
above in the order of the (properly defined) increasing perturbation strength,
constitute the framework for classification of LE decay regimes
\cite{gorin,jacq-2}. It is important to mention that the full variety of
system- and perturbation-dependent decay regimes is rather rich, and extends
far beyond the above list: double-Lyapunov \cite{wang}, super-exponential
\cite{silv} and power law \cite{Mirlin} decay regimes serve as examples. We
further note that analytical results for the time decay of the {\it
  unaveraged} LE, $M(t)$, are currently available only for very few chaotic
systems \cite{iomin}.

The discovery \cite{jalab} of the Lyapunov regime for the decay of the
averaged LE in classically chaotic systems, $\overline{M(t)} \sim
\exp(-\lambda t)$ with $\lambda$ being the average Lyapunov exponent, provided
a strong and appealing connection between classical and quantum chaos: it
related a measure of instability of the quantum dynamics, such as the LE, to a
quantity characterizing the instability of the corresponding classical
dynamics, i.e. the Lyapunov exponent. This result awoke the interest on the LE
in the quantum chaos community. The Lyapunov regime has been numerically
observed in several two-dimensional chaotic systems, i.e. in the Lorentz gas
\cite{cucch-1,cucch-2}, the Bunimovich stadium \cite{wisn}, the smooth stadium
billiard \cite{cucch-3}, a Josephson flux qubit device \cite{dominguez}, as
well as in one-dimensional time-dependent Hamiltonian systems \cite{jacq-1}.

The theory of the Lyapunov decay of the LE mainly relies on the following two
assumptions: (i) the validity of the structural stability arguments (supported
by the shadowing theorem \cite{vanic}), and (ii) the global nature of the
Hamiltonian perturbation. The first assumption guarantees a unique one-to-one
mapping of trajectories of the unperturbed system to those of the perturbed
system. This mapping allows for efficient pairing of the trajectories of the
unperturbed and perturbed system in the diagonal approximation \cite{berry}.
The second assumption implies that the Hamiltonian perturbation affects every
trajectory of the system, and, therefore, all trajectories are responsible for
the decay of the LE. However, this is by no means the most general situation
when we consider possible experimental realizations of the LE.

In the present work we extend the semiclassical theory of the LE by lifting
the second of the two above-mentioned assumptions, i.e. {\it we allow for a
  local perturbation in coordinate space}. In this context the LE decay was
previously addressed in the case of a {\it strong} local perturbation
\cite{gouss}, i.e. for a billiard exposed to a local boundary deformation much
larger than the de Broglie wavelength. Analytical and numerical calculations
yielded a novel LE decay regime, for which $M(t) \sim \exp(-2\gamma t)$ with
$\gamma$ being the probability (per unit time) of the corresponding classical
particle to encounter the boundary deformation. $\gamma$ can also be viewed as
a classical {\it escape rate} from a related open billiard obtained from the
original (closed) one by removing the deformation-affected boundary segment.
In this work we explore all possible strengths of a local perturbation and
describe the transition from the weak to the strong perturbation regime that
completes the previous picture. In particular we show that the rate of the
exponential decay of the LE oscillates as a function of the perturbation
strength as the escape-rate regime is approached.

The paper is organized as follows: In Section~\ref{section_theory} we develop
a comprehensive semiclassical approach of the LE decay due to local
Hamiltonian perturbations of increasing strength. We perform a systematic
analysis of the different decay regimes and establish their relation to the
previously known decay regimes in the case of a global perturbation. In
Section~\ref{section_numerics} we validate our semiclassical theory by
numerical simulations. In Sec.~\ref{section_experiment} we outline possible
experimental realizations and focus on the possibility of introducing a local
perturbation in microwave-cavities and ultra-cold atom-optics billiards. We
provide concluding remarks in Sec.~\ref{section_conclusion} and point to the
similarities and differences with respect to another approach to the LE for
local perturbations. Technical aspects of the calculations are relegated to
the appendices.

\section{Semiclassical approach}
\label{section_theory}

\subsection{Wave-function evolution}

We address the time evolution of the wave function that
describes a quantum particle moving inside a classically chaotic
two-dimensional billiard (corresponding to a Hamiltonian $H$). We assume
that initially (at time $t=0$) the particle is in a coherent 
state
\begin{equation}
  \phi_0(\vr) = \frac{1}{\sqrt{\pi}\sigma} \ \exp\left[
    \frac{i}{\hbar}\vp_0\cdot(\vr-\vr_0) -
    \frac{(\vr-\vr_0)^2}{2\sigma^2} \right]
  \label{S1}.
\end{equation}
Here $\sigma$ quantifies the extension of the Gaussian wave packet, while
$\vr_0$ and $\vp_0$ are the initial mean values of the position and momentum
operators, respectively. We further define the (rescaled \cite{footnote-1}) de
Broglie wavelength of the particle as $\lambdabar = \hbar/p_0$.

In our description of the time evolution of the wave function we rely
on the semiclassical approximation \cite{brack} of the wave function 
at a time $t$,
\begin{equation}
  \phi_t(\vr) = \int d\vr' \sum_{\hat{s}(\vr,\vr',t)}
  \K_{\hat{s}}(\vr,\vr',t) \phi_0(\vr') \; .
  \label{S2}
\end{equation}
Here
\begin{equation}
  \K_{\hat{s}}(\vr,\vr',t) = \frac{\sqrt{D_{\hat{s}}}}{2\pi i\hbar}
  \ \exp\left[ \frac{i}{\hbar}S_{\hat{s}}(\vr,\vr',t) -
    i\frac{\pi\nu_{\hat{s}}}{2} \right]
  \label{S3}
\end{equation}
is the contribution to the Van Vleck propagator associated with the 
classical trajectory $\hat{s}(\vr,\vr',t)$ leading from point $\vr'$ 
to point $\vr$ in time $t$. $S_{\hat{s}}(\vr,\vr',t)$ 
denotes the classical action integral (or the Hamilton principal 
function) along the path $\hat{s}$. In a hard-wall billiard 
$S_{\hat{s}}(\vr,\vr',t) = (m/2t) L_{\hat{s}}^2(\vr,\vr')$, where 
$L_{\hat{s}}(\vr,\vr')$ is the length of the trajectory $\hat{s}$, and 
$m$ is the mass of the particle. In Eq.~(\ref{S3}), $D_{\hat{s}} =
|\det(-\partial^2S_{\hat{s}} / \partial\vr \partial\vr')|$, and the Maslov 
index $\nu_{\hat{s}}$ equals the number of caustics along the trajectory 
$\hat{s}$ plus twice the number of particle-wall collisions (for
the case of Dirichlet boundary conditions).

\begin{figure}[h]
\centerline{\epsfig{figure=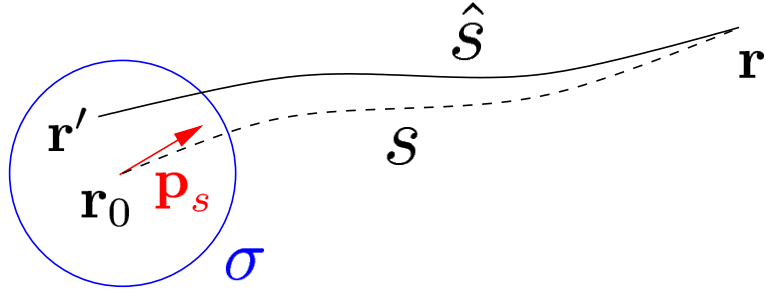,width=4.0in}}
\caption{Sketch of a typical trajectory $\hat{s}$ (full line), connecting a
  point $\vr'$ (within the circular extension of radius $\sigma$ of the
  initial wave packet) to the point $\vr$, where the evolved wave function is
  evaluated, together with the central trajectory $s$ (dashed line) that
  reaches the same final point, but starts at the center $\vr_0$ of the wave
  packet (with momentum $\vp_s$). The linearization of Eq.~(\ref{S4}),
  together with the conditions discussed in the text, allow to represent all
  the trajectories $\hat{s}$ contributing to Eq.~(\ref{S2}) by the single
  reference trajectory $s$.}
\label{fig-1}
\end{figure}
Since we assume that the initial wave packet is localized around $\vr_0$
within $\sigma$, only trajectories starting at points $\vr'$ close to $\vr_0$
are relevant for our semiclassical description.  Thus we can expand the action
integral $S_{\hat{s}}(\vr,\vr',t)$ in a power series in $(\vr'-\vr_0)$. In
\ref{app_action} we show that the power series can be terminated at the linear
term,
\begin{equation}
  S_{\hat{s}}(\vr,\vr',t) \approx S_s(\vr,\vr_0,t) -
  \vp_s\cdot(\vr'-\vr_0) \; ,
  \label{S4}
\end{equation}
if the wave packet is narrow enough, so that the condition
\begin{equation}
  \sigma \ll \sqrt{\frac{\lL}{1/\lambdabar+1/\sigma}}
  \label{condition-1}
\end{equation}
is satisfied, where $\lL = p_0/m\lambda$ is the Lyapunov length.
In Eq.~(\ref{S4}),  
$s(\vr,\vr_0,t)$ is the {\it central} reference trajectory into which 
${\hat{s}}(\vr,\vr',t)$ gets uniformly deformed as 
$\vr' \rightarrow \vr_0$, and $\vp_s = -\partial S_s(\vr,\vr_0,t) /
\partial\vr_0$ denotes the initial momentum 
of the trajectory $s$ (see Fig.~\ref{fig-1}).

Substituting Eq.~(\ref{S4}) into Eq.~(\ref{S3}), and performing
the integration in Eq.~(\ref{S2}) we obtain 
\begin{equation}
  \phi_t(\vr) = 2\pi\hbar \sum_{s(\vr,\vr_0,t)} \K_s(\vr,\vr_0,t) \ 
  \Phi_0(\vp_s) \ , 
  \label{S5}
\end{equation}
with 
\begin{eqnarray}
  \Phi_0(\vp) &\equiv& \int \frac{d\vr}{2\pi\hbar} \exp\left[
    -\frac{i}{\hbar}\vp\cdot(\vr-\vr_0) \right] \phi_0(\vr)
  \label{S6}\\
  &=& \frac{\sigma}{\sqrt{\pi}\hbar} \exp\left[
    -\frac{\sigma^2}{2\hbar^2}(\vp-\vp_0)^2 \right]
  \nonumber
\end{eqnarray}
the momentum representation of the initial wave packet.

\begin{figure}[h]
\centerline{\epsfig{figure=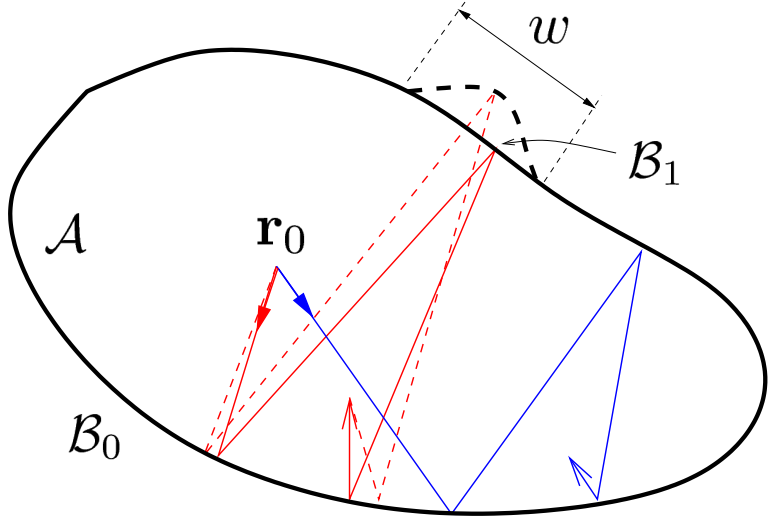,width=4.0in}}
\caption{Sketch of our model system of a particle moving inside a chaotic
  billiard of area $\A$. The perturbation consists of a deformation localized
  in a region ${\cal B}_1$ (of width $w$) of the billiard boundary. The
  complementary set ${\cal B}_0$ of the boundary is unaffected by the
  perturbation. The two trajectories $s$ (red solid line) and $\hat{s}$ (red
  dashed line) starting from $\vr_0$ with different momenta correspond,
  respectively, to the unperturbed and perturbed Hamiltonian. The diagonal
  approximation entering Eq.~(\ref{S9}) identifies both of them and assigns an
  action difference given by Eq.~(\ref{S11}). The starting momentum of the
  solid red trajectory belongs to the set $\P_1$.  The third trajectory (blue
  solid line) hits the boundary only at ${\cal B}_0$ and therefore is the same
  for both the unperturbed and perturbed systems. Hence the action difference
  of the corresponding trajectory pair is zero. The starting momentum of the
  blue trajectory belongs to the set $\P_0$.}
\label{fig-2}
\end{figure}

We now consider a related billiard, corresponding to the perturbed
Hamiltonian $\tilde{H}$, that differs from the original (unperturbed)
billiard by a deformation of the boundary segment ${\cal B}_1$ of width
$w$ (see Fig.~\ref{fig-2}). The perturbation is thus local, and 
will be characterized by its extent (depending on the ratio between
$w$ and the cavity perimeter $P$) and its strength (that will be
quantified in the sequel). In view of Eq.~(\ref{S5}), the wave function 
describing the evolution of the particle (starting from the same initial 
state $\phi_0$) can be written as
\begin{equation}
  \tilde{\phi}_t(\vr) = 2\pi\hbar \sum_{\tilde{s}(\vr,\vr_0,t)}
  \K_{\tilde{s}}(\vr,\vr_0,t) \Phi_0(\vp_{\tilde{s}}) \, .
  \label{S7}
\end{equation}
The sum now runs over all possible trajectories
$\tilde{s}(\vr,\vr_0,t)$ of a classical particle that travels from
$\vr_0$ to $\vr$ in time $t$ while bouncing off the boundary of the
perturbed billiard.

\subsection{Wave-function overlap for local perturbations}
\label{section_wfoflp}

According to Eqs.~(\ref{S5})-(\ref{S7}) and the definition (\ref{I1}) 
of the LE amplitude, we have 
\begin{eqnarray}
  O(t) &=& \int_{\A}d\vr \ \tilde{\phi}_t^*(\vr) \phi_t(\vr)
  \label{S8}\\
  &=& \int_{\A}d\vr
  \sum_{\tilde{s}} \sum_s \sqrt{D_{\tilde{s}}D_s}
  \nonumber\\
  &&\times \exp\left[
    \frac{i}{\hbar}(S_s-S_{\tilde{s}}) - i\frac{\pi(\nu_s-\nu_{\tilde{s}})}{2}
  \right] \Phi_0^*(\vp_{\tilde{s}})\Phi_0(\vp_s) \, ,
  \nonumber
\end{eqnarray}
where ${\A}$ stands for the billiard area. The shadowing theorem \cite{vanic}
allows us to employ the diagonal approximation ($s\simeq \tilde{s}$) in the
case of a classically small perturbation \cite{footnote-2}, thus reducing
Eq.~(\ref{S8}) to
\begin{equation}
  O(t) = \int_{\A}d\vr \sum_{s(\vr,\vr_0,t)} D_s \ \exp\left[
    \frac{i}{\hbar}\Delta S_s(\vr,\vr_0,t) \right] W_0(\vp_s) \, ,
  \label{S9}
\end{equation}
where
\begin{equation}
  W_0(\vp) \equiv \left| \Phi_0(\vp) \right|^2 =
  \frac{\sigma^2}{\pi\hbar^2} \ \exp\left[
    -\frac{\sigma^2}{\hbar^2}(\vp-\vp_0)^2 \right]
  \label{S10}
\end{equation}
is the probability distribution of the particle momentum. In
billiards the action difference between the two trajectories
traveling between the same initial and final points in the same time 
$t$ can be written, in terms of their length difference $\Delta L_s$, as
\begin{eqnarray}
  \Delta S_s(\vr,\vr_0,t) &\equiv& S_s - S_{\tilde{s}}
  \label{S11}\\
  &=& \frac{p_s^2}{2m}t -
  \frac{p_{\tilde{s}}^2}{2m}t \approx p_s\frac{p_s-p_{\tilde{s}}}{m}t
  = p_s \ \Delta L_s(\vr,\vr_0,t) \, .
  \nonumber
\end{eqnarray}
Using the Jacobian property of the Van Vleck determinant $D_s$, we
can replace the integral over final coordinates in Eq.~(\ref{S9}) 
by an integral over the initial momenta and obtain
\begin{equation}
  O(t) = \int d\vp \, W_0(\vp) \exp\left[ \frac{i}{\hbar} \ p \ \Delta
    L(\vr_0,\vp t) \right] \, .
  \label{S12}
\end{equation}
The dependence of $\Delta L$ on the product $\vp t$ stems from the
fact that in billiards, changing the magnitude of the momentum only 
modifies the traveling time, but does not affect the path.

We now introduce a sequence of momentum sets $\P_n(\vr_0,t)$, with
$n=0,1,2,\ldots$, such that for any $\vp \in \P_n$ the classical trajectory,
starting from the phase-space point $(\vr_0,\vp)$, arrives at a
coordinate-space point $\vr \in \A$ after time $t$ while visiting the
deformation-affected boundary segment (${\cal B}_1$ in Fig.~\ref{fig-2})
exactly $n$ times. Thus, the trajectories with the initial momentum in
$\P_0(\vr_0,t)$ only undergo collisions with the part of the boundary
unaffected by the deformation (${\cal B}_0$ in Fig.~\ref{fig-2}) rendering
$\Delta L=0$, so that
\begin{equation}
  O(t) = \sum_{n=0}^\infty O_n(t) \, , 
  \label{eq-O-decomposition}
\end{equation}
with 
\begin{eqnarray}
  O_0(t) & = &  \int_{\P_0}d\vp \ W_0(\vp) \, ,
  \label{eq-O_0}\\
  O_n(t) & = &  \int_{\P_n}d\vp \ W_0(\vp) \ e^{ip\Delta L(\vr_0,\vp t)/\hbar}
  \;\;\;\;\; \mathrm{for} \;\;\;\;\; n\ge 1 \, .
  \label{eq-O_1}
\end{eqnarray}

Since we are studying billiards, the integrations over momenta can be
simplified by working in polar coordinates $(p,\theta)$ and considering the
sets $\Theta_n(\vr_0,pt)$ of angles $\theta$ such that $\vp\equiv (p,\theta)
\in {\P_n}(\vr_0,t)$ iff $\theta \in \Theta_n$. For a classically chaotic
dynamics the set $\Theta_0$ shrinks with increasing time $t$, and $\Theta_0$
becomes the fractal set defining the repeller of the corresponding open
(scattering) problem in the limit $t\rightarrow \infty$. Eqs.~(\ref{eq-O_0})
and (\ref{eq-O_1}) can, respectively, be written as
\begin{eqnarray}
  O_0(t) & = & 
  \int_0^\infty dpp \int_{\Theta_0}d\theta \ W_0(\vp) \, , 
  \label{eq-O_0-b}\\
  O_n(t) & = &
  \int_0^\infty dpp \int_{\Theta_n}d\theta \ W_0(\vp) \ e^{ip\Delta
    L(\vr_0,p t,\theta)/\hbar} \;\;\;\;\; \mathrm{for} \;\;\;\;\; n\ge 1 \, .
  \label{eq-O_1-b}
\end{eqnarray}
For long times $t$, where many trajectories contribute to the semiclassical
expansions, the angular integrals over $\Theta_n$ can be replaced by integrals
over all angles, weighted with the measures $\rho_n$ of the corresponding
sets. Assuming ergodicity one readily obtains an approximation for the
probability for a trajectory of length $l=pt/m$ to visit the boundary region
${\cal B}_1$ exactly $n$ times:
\begin{equation}
  \rho_n(l) \approx \frac{1}{n!} \left( \frac{l}{\lD} \right)^n 
  \exp\left( -\frac{l}{\lD} \right) \, ,
  \label{eq-rho}
\end{equation}
where $\lD$ is the average dwell length of paths in the related open chaotic
billiard obtained from the original (closed) one by removing the boundary
region ${\cal B}_1$. This corresponds to the classical escape rate of the open
cavity that for particles with momentum $p_0$ is given by
\begin{equation}
  \gamma = \frac{p_0}{m\lD} \, .
  \label{eq-escape-rate}
\end{equation}
For a chaotic cavity with an opening $w$ much smaller than its perimeter
$P$ we can approximate \cite{bauer} $\lD \approx \pi \A/w$, and therefore
\begin{equation}
  \gamma \approx \frac{p_0}{m} \ \frac{w}{\pi \A} \ . 
  \label{S41}
\end{equation}
In our case the escape rate $\gamma$ yields a measure of the perturbation
extent. The classical escape rate of an open cavity controls the fluctuations
of the transmission coefficients, and therefore approximations such as
(\ref{S41}) have been thoroughly examined in the context of quantum
transport \cite{chaos_93}.

According to the previous discussion we can approximate $O_n(t)$, with
$n=0,1,2,\ldots$, by the averages
\begin{eqnarray}
\bar{O}_0(t) & = &   \int_0^\infty dp \ p \: \rho_0(pt/m)
  \int_0^{2\pi}d\theta \ W_0(\vp)  \, ,
  \label{eq-O_0-bar}\\
  \bar{O}_n(t) & = &
  \int_0^\infty dp \ p \left\langle e^{ip\Delta L(\vr_0,p t,\theta)/\hbar}
  \right\rangle_{\Theta_n} \rho_n(pt/m)
  \int_0^{2\pi}d\theta \ W_0(\vp) \; , \; n\ge 1.
  \label{eq-O_1-bar}
\end{eqnarray}
The mean value $\langle \: \ldots \: \rangle_{\Theta_n}$ should be taken over
the set $\Theta_n(\vr_0,p t)$. The chaotic nature of the dynamics will enable
us to treat the averages over $\Theta_n$ in a statistical way. In view of
Eq.~(\ref{S10}), the $\theta$-integral in Eqs.~(\ref{eq-O_0-bar}) and
(\ref{eq-O_1-bar}) yields
\begin{equation}
  \int_0^{2\pi}d\theta W_0(\vp) = \frac{2\sigma^2}{\hbar^2} \exp\left[
    -\frac{\sigma^2}{\hbar^2}\left(p^2+p_0^2\right) \right] I_0\left(
    \frac{2\sigma^2}{\hbar^2}p_0p \right),
  \label{S20}
\end{equation}
where $I_0$ denotes the modified Bessel function. 

As usually assumed in the Loschmidt echo studies, we restrict our
analysis to ``semiclassical'' initial wave packets $\phi_0(\vr)$
with sizes much larger than the de Broglie wave length,
\begin{equation}
  \lambdabar \ll \sigma \, .
  \label{condition-2}
\end{equation}
This assumption, together with condition 
(\ref{condition-1}), defines the interval for the
dispersion $\sigma$, where the semiclassical approach is reliable,
and hence yields restrictions to the parameters of the
billiard. Employing condition (\ref{condition-2})  enables us to use
the asymptotic form $I_0(x) \approx e^x/\sqrt{2\pi x}$, valid for large $x$,
in Eq.~(\ref{S20}). Thus, the probability distribution function for the
magnitude of the initial momentum is given by
\begin{equation}
  p\int_0^{2\pi}d\theta W_0(\vp) \approx \frac{\sigma}{\hbar}
  \sqrt{\frac{p}{\pi p_0}} \ \exp\left[
    -\frac{\sigma^2}{\hbar^2}(p-p_0)^2 \right].
  \label{S21}
\end{equation}
We note that Eq.~(\ref{S21}) provides a good approximation to the exact
distribution function already for $\sigma \gtrsim 2\lambdabar$. Under this
assumption the $p$-integrals in Eqs.~(\ref{eq-O_0-bar}) and (\ref{eq-O_1-bar})
are dominated by the contributions around $p_0$, and we can write (in view of
Eq.~(\ref{eq-rho}))
\begin{eqnarray}
  \bar{O}_0(t) &\approx& e^{-\gamma t} \, ,
  \label{S22-n_eq_0}\\
  \bar{O}_n(t) &\approx& \frac{(\gamma t)^n}{n!} e^{-\gamma t} \left\langle
    e^{i p_0 \Delta L(\vr_0,p_0 t,\theta)/\hbar} \right\rangle_{\Theta_n}
  \;\;\;\;\; \mathrm{for} \;\;\;\;\; n\ge 1 \, .
  \label{S22}
\end{eqnarray}
Since in Eq.~(\ref{S22}) all classical quantities are evaluated for an
initial momentum with magnitude $p_0$, the mean values $\langle \: \ldots \:
\rangle_{\Theta_n}$ should be taken over the sets $\Theta_n(\vr_0,p_0 t)$.
However, for long times and a chaotic dynamics we do not expect these mean
values to depend on $\vr_0$.  In the next section we will further invoke the
chaotic nature of the underlying classical dynamics in order to estimate the
mean values and therefore the LE average amplitude.

\subsection{Averages over trajectory distributions}
\label{subsec-averages}

For classically small perturbations the action, respectively, length difference 
 (Eq.~(\ref{S11})) between a 
trajectory $s$ (solid red segment in Fig.~\ref{fig-2}) and its perturbed 
partner $\tilde{s}$ (dashed segment) is given  only by 
the contributions accumulated along the $n$ encounters with ${\cal B}_1$.
Differences in length arising from the free flights between collisions
with the boundary (${\cal B}_0+{\cal B}_1$) are of higher order
in the perturbation strength and will not be considered. We can
then write
\begin{equation}
  \Delta L = \sum_{j=1}^n u(\vartheta_j, \xi_j) \, ,
  \label{S24}
\end{equation}
where the deformation function $u(\vartheta_j, \xi_j)$ is the length
difference accumulated in the $j$-th collision with ${\cal B}_1$, depending on
the impinging angle $\vartheta_j \in (-\pi/2,\pi/2)$ and on the coordinate
$\xi_j \in (0,w)$ of the hitting point. The number $n$ of collisions with
${\cal B}_1$ is a fraction of the total number of collisions $p_0t/m\lF$. The
mean bouncing length $\lF$ can be approximated by $\pi \A/P$ \cite{niels}, and
we suppose $\lD \gg \lF$ since $P \gg w$. We note that for small perturbations
$\Delta L$ depends on $p$ only through $n$.

Given the chaotic nature of the classical dynamics and the fact that the
collisions with ${\cal B}_1$ are typically separated by many collisions with
${\cal B}_0$ we assume $\{\vartheta_j\}$ and $\{\xi_j\}$ to be random
variables.  Furthermore, assuming a perfect randomization of the trajectories
within the billiard the probability distribution functions for
$\{\vartheta_j\}$ and $\{\xi_j\}$ are, respectively,
\begin{equation}
  P_\vartheta(\vartheta) = \frac{\cos\vartheta}{2} \qquad 
  \mathrm{and} \qquad P_\xi(\xi) = \frac{1}{w} \, .
  \label{S27}
\end{equation}
Then, treating the random variables as uncorrelated we obtain
\begin{equation}
  \left\langle e^{i p_0 \Delta L(\vr_0,p_0 t,\theta)/\hbar}
  \right\rangle_{\Theta_n} = \left\langle e^{i u / \lambdabar} \right\rangle^n
  \, ,
  \label{S28-new}
\end{equation}
where the average $\langle \ldots \rangle$ is defined as
\begin{equation}
  \left\langle f \right\rangle \equiv \int_{-\pi/2}^{\pi/2} d\vartheta P_\vartheta
  (\vartheta) \int_0^w d\xi P_\xi(\xi) f(\vartheta,\xi)
  \label{S29-new}
\end{equation}
for a function $f(\vartheta,\xi)$. Once we specify the shape of the
perturbation, the average in the right-hand side of Eq.~(\ref{S28-new}) is
readily calculated from the probability distributions of Eq.~(\ref{S27}). For
instance, for a piston-like deformation (see \ref{app_piston}),
\begin{equation}
  u(\vartheta,\xi)=2h\cos\vartheta \, ,
  \label{eq:piston}
\end{equation}
so that the average reads
\begin{equation}
  \left\langle e^{i u / \lambdabar} \right\rangle = 1 -
  \frac{\pi}{2} \left[ {\bf H}_1(2h/\lambdabar) - 
  i J_1(2h/\lambdabar) \right] \, ,
  \label{piston-average}
\end{equation}
where ${\bf H}_1$ stands for the Struve H-function of the first order, and
$J_1$ is the first order Bessel function of the first kind. Here we note that
for the case of the piston-like boundary deformation the ratio $h/\lambdabar$
serves as a measure of the perturbation strength, whereas $w$ (and therefore
$\gamma$) quantifies the extent of the perturbation.

At this stage, however, we will keep our discussion general and do not specify
the details of the local perturbation. The substitution of Eq.~(\ref{S28-new})
into Eq.~(\ref{S22}) yields the average LE amplitude. The latter is usually
not an observable quantity, however it will be helpful towards our
semiclassical calculation of the LE. The condition of a classically small
perturbation that we have adopted throughout our work, implies that
$\sqrt{\langle u^2\rangle} \ll w \ll P$ (which for the case of the piston-like
deformation is equivalent to $h \ll w \ll P$). Quantum mechanically the
perturbation can be characterized by the extent $w$ and a deformation strength
\begin{equation}
  \chi \equiv \frac{\sqrt{\langle u^2\rangle}}{\lambdabar} \, .
  \label{chi-def}
\end{equation}
For $\chi \ll 1$ we will be in the quantum perturbative regime
\cite{jacq-1,toms,pros}, which will not be considered in this work. Increasing
the deformation strength $\chi$ we anticipate a richer variety of regimes than
for the case of the LE under global perturbations \cite{jalab,gorin} since the
perturbation extent enters as another relevant parameter.

\subsection{Loschmidt echo for local perturbations}

According to Eqs.~(\ref{LE}) and (\ref{S8}) the semiclassical expansion
for the LE contains terms involving four trajectories. The diagonal
approximation, leading to Eq.~(\ref{S9}) for the LE amplitude, reduces
the LE to a sum over pairs of trajectories. Consequently, the semiclassical 
form of the LE must take into account the different possibilities
for each trajectory of the pair to hit (or not) the region of the 
boundary where the perturbation acts. We can therefore decompose the
LE as
\begin{equation}
  M(t) =  M^{\mathrm{nd}}(t) + M^{\mathrm{d}}(t) \, ,
  \label{S14}
\end{equation}
where we have introduced the non-diagonal and diagonal contributions
according to
\begin{equation}
  M^{\mathrm{nd}}(t) =\int d\vp W_0(\vp)
  e^{-ip\Delta L(\vr_0,\vp t)/\hbar}
  \int_{\bar{\varepsilon}_\vp}d\vp' W_0(\vp') e^{ip'\Delta
    L(\vr_0,\vp' t)/\hbar}
  \label{S15}
\end{equation}
and
\begin{eqnarray}
  M^{\mathrm{d}}(t) &=& \int d\vp W_0(\vp)
  \int_{\varepsilon_\vp}d\vp' W_0(\vp')
  \label{S16}\\
  &&\times \exp\left\{ \frac{i}{\hbar}
    \left[ p'\Delta L(\vr_0,\vp' t)-p\Delta L(\vr_0,\vp t) \right]
  \right\}.
  \nonumber
\end{eqnarray}
Here, the set $\varepsilon_\vp$ of momenta $\vp'$ is defined such that two
trajectories starting from the phase space points $(\vr_0,\vp)$ and
$(\vr_0,\vp')$, stay ``close'' to each other in phase space during time $t$,
and thus are ``correlated'' with respect to the perturbation;
$\bar{\varepsilon}_\vp$ is the momentum set complementary to
$\varepsilon_\vp$. We give a quantitative definition of $\varepsilon_\vp$
below. Following the standard notation introduced in Ref.~\cite{jalab}, we
call diagonal term the one resulting from the identification of pairs of
trajectories where the effect of the perturbation is correlated, that is, when
$\vp' \in \varepsilon_\vp$.  In the non-diagonal term we consider the pairs of
trajectories uncorrelated with respect to the perturbation, including the case
where one or both orbits are unperturbed. As noted at the beginning of this
section, each of the trajectories of the above pair already incorporates a
diagonal approximation between a perturbed and an unperturbed trajectories
with the same extreme points.

\subsection{Non-diagonal contribution to the Loschmidt echo}

Calculating the LE as an average over trajectory distributions forces us to
take into account pairs of trajectories and the possible correlations among
them. The correlations are particularly important for $M^\mathrm{d}(t)$, as we
show in Sec.~\ref{section_diagonal}. On the other hand, in our discussion of
the last chapter we established that for the calculation of $M^\mathrm{nd}(t)$
the two trajectories of the pair can be considered to be uncorrelated with
respect to the perturbation, and the averages can be performed independently.
Assuming in addition that the measure of the momentum set $\varepsilon_\vp$ is
small compared with that of the momentum set effectively represented by the
distribution $W_0(\vp)$ we write
\begin{equation}
  M^{\mathrm{nd}}(t) \approx \left| \sum_{n=0}^\infty \bar{O}_n(t) \right|^2.
  \label{S19}
\end{equation}
Substituting Eqs.~(\ref{S22-n_eq_0}-\ref{S22}) and (\ref{S28-new}) into
Eq.~(\ref{S19}) and performing straightforward algebraic operations we arrive
at our central result for the non-diagonal contribution to the LE:
\begin{equation}
  M^\mathrm{nd}(t) \approx e^{- \kappa \gamma t} \, ,
  \label{S32}
\end{equation}
where
\begin{equation}
  \kappa \equiv 2 \left( 1-\Re \left\langle
      e^{i u / \lambdabar} \right\rangle \right) \, .
  \label{kappa}
\end{equation}
Thus, for the case of the piston-like boundary deformation the average phase
factor due to a single visit of the perturbation region ${\cal B}_1$ by the
particle is given by Eq.~(\ref{piston-average}), so that Eq.~(\ref{kappa})
reads
\begin{equation}
  \kappa = \pi {\bf H}_1(2h/\lambdabar) \, .
%  M^\mathrm{nd}(t) \approx \exp\left[ -\pi {\bf H}_1(2h/\lambdabar) \gamma t \right].
  \label{S32-piston}
\end{equation}

In subsection \ref{subsec_decay} we study the emergence of the
Fermi-golden-rule and escape-rate regimes of the decay of $M^\mathrm{nd}$ (as
well as the transition between the two regimes) depending on the strength of
the perturbation. In sections~\ref{section_numerics} and
\ref{section_experiment} we discuss their implications for numerical
simulations and possible experimental observations.

\subsection{Diagonal contribution to the Loschmidt echo}
\label{section_diagonal}

To proceed with the calculation of the diagonal contribution to the
LE, Eq.~(\ref{S16}), we first need to specify the set
$\varepsilon_\vp$ such that two trajectories of time $t$, starting
from the phase space points $(\vr_0,\vp)$ and
$(\vr_0,\vp'\in\varepsilon_\vp)$, stay ``correlated'' during time
$t$. As in Sec.~\ref{section_wfoflp}, it is convenient to work with
polar coordinates, $\vp=(p,\theta)$ and $\vp'=(p',\theta')$, in which 
the set $\varepsilon_\vp$ can be defined as follows: for every 
$\vp' \in \varepsilon_\vp$ one has $|p'-p| \lesssim \Delta p$ and
$|\theta'-\theta| \lesssim \Delta\theta$. In turn, $\Delta p$ and
$\Delta \theta$ are subject to the requirement that the two
trajectories stay ``close'' to each other in phase space. Indeed,
any two ``correlated'' trajectories must have the same number of
collisions with the billiard boundary. This condition leads to
$|p'-p|t/m \lesssim \lF$. Moreover, the two trajectories must also 
have the same number of collisions with ${\cal B}_1$. The
spatial separation between the two trajectories at the first collision
with the boundary is approximately given by $|\theta'-\theta|
\lF$. The condition that after time $t$ this separation is smaller
that the size $w$ of ${\cal B}_1$ is 
$|\theta'-\theta| \lF \exp(\lambda t) \lesssim w$. Here
we have used the property that for chaotic dynamics two arbitrary,
initially close trajectories deviate exponentially from each
over with a rate given by the average Lyapunov exponent $\lambda$.
 Thus, we can estimate the measure of the $\varepsilon_\vp$-set to be
\begin{equation}
  \Delta p = \frac{m \lF}{t} \;\;\;\;\; \mathrm{and} \;\;\;\;\;
  \Delta\theta = \frac{w}{\lF} \exp(-\lambda t) \, .
  \label{S33}
\end{equation}
Using this quantitative description of $\varepsilon_\vp$ for the evaluation
of Eq.~(\ref{S16}) we obtain for the diagonal contribution
to the LE
\begin{eqnarray}
  M^{\mathrm{d}}(t) = \int d\vp W_0(\vp) &\int_{p-\Delta
    p}^{p+\Delta p}dp' p' \,\,
  \int_{\theta-\Delta\theta}^{\theta+\Delta\theta} d\theta'
  W_0(\vp')
  \label{S34}\\
  &\times \exp\left\{ \frac{i}{\hbar} \left[ p'\Delta
      L(\vr_0,p't,\theta')-p\Delta L(\vr_0,pt,\theta) \right] \right\}.
  \nonumber
\end{eqnarray}
We now argue that for boundary deformations of moderate strength the
exponent in the integrand on the right hand side of Eq.~(\ref{S34})
can be neglected. The argument of the exponent is given by the total
differential of the function $p\Delta L(\vr_0,pt,\theta)$, and
therefore its absolute value can be bounded by
\begin{equation}
  \left| \frac{(p'-p)\Delta L}{\hbar} +
  \frac{p(\theta'-\theta)}{\hbar}\frac{\partial\Delta
  L}{\partial\theta} \right| \lesssim \left| \frac{\Delta p\Delta
  L}{\hbar} \right| + \left|
  \frac{p\Delta\theta}{\hbar}\frac{\partial\Delta L}{\partial\theta}
  \right|.
  \label{S34.1}
\end{equation}
Here we have used that, as discussed in Sec.~\ref{subsec-averages}, $\Delta L$
is independent of $p$ for a fixed number $n$ ($\approx \gamma t$) of
collisions with ${\cal B}_1$. Then
\begin{equation}
  \left| \frac{\Delta p\Delta L}{\hbar} \right| \lesssim \frac{\Delta p}{\hbar}
  \sqrt{\langle u^2\rangle} \gamma t = \frac{\lF}{\lD}\chi \, ,
  \label{S34.2}
\end{equation}
and
\begin{equation}
  \frac{p\Delta\theta}{\hbar}\frac{\partial\Delta L}{\partial\theta}
  = \frac{p\Delta\theta}{\hbar} \sum_{j=1}^n \left( \frac{\partial
      u}{\partial\vartheta_j}\frac{\partial\vartheta_j}{\partial\theta} +
    \frac{\partial u}{\partial\xi_j}\frac{\partial\xi_j}{\partial\theta}
  \right) \sim \Delta\theta \, C e^{\lambda t} =
  \frac{w}{\lF} C\, ,
  \label{S34.3}
\end{equation}
where we have introduced the dimensionless quantity
\begin{equation}
  C = \frac{1}{\lambdabar} \left\langle \frac{\partial u}{\partial\vartheta}
  \right\rangle + \frac{\lF}{\lambdabar} \left\langle \frac{\partial u}{\partial\xi}
  \right\rangle.
  \label{S34.4}
\end{equation}
This implies that the exponent in Eq.~(\ref{S34}) is smaller than unity if
$\chi \lesssim \lD/\lF$ and $C \lesssim \lF/w$. Since the ratios $\lD/\lF$ and
$\lF/w$ are assumed to be large, the above inequalities hold for a wide range
of deformations. (Note that for the piston-like deformation,
Eq.~(\ref{eq:piston}), $\chi= (8/3)^{1/2} \, h / \lambdabar$ and $C=0$, and
hence the exponent is small if $h \lesssim \lambdabar \lD/\lF$.)

Neglecting the exponent in Eq.~(\ref{S34}) we obtain
\begin{eqnarray}
  M^{\mathrm{d}}(t) &\approx& \int d\vp W_0(\vp) \int_{p-\Delta
    p}^{p+\Delta p}dp' p'
  \int_{\theta-\Delta\theta}^{\theta+\Delta\theta} d\theta'
  W_0(\vp')
  \label{S35}\\
  &\approx& 4\Delta p \Delta\theta \int d\vp p W_0^2(\vp) \, .
  \nonumber
\end{eqnarray}
In the second line of this equation we have taken into account that 
$\Delta p \ll p$ and $\Delta\theta \ll 1$ for times $t$ much longer 
than the dwell time $t_\mathrm{d}$. Under the assumption
(\ref{condition-2}) of a ``semiclassical'' initial wave packet, the
integral over $p$ in Eq.~(\ref{S35}) is dominated by the contribution
around $p_0$ and we find
\begin{equation}
  M^{\mathrm{d}}(t) \approx \frac{2 m w}{\pi p_0 t} \left(
    \frac{\sigma}{\lambdabar} \right)^2 e^{-\lambda t}.
  \label{S36}
\end{equation}

We note that the exponential dependence of $M^{\mathrm{d}}$ on $\lambda t$
does not contain the perturbation and arises from a classical probability
distribution, as in the standard Lyapunov regime \cite{jalab}. The dependence
with respect to the perturbation extent, $w$, appears in the prefactor.

\subsection{Decay regimes of the Loschmidt echo}
\label{subsec_decay}

According to Eq.~(\ref{S14}) the full LE $M(t)$ is the sum of the non-diagonal
and diagonal contributions, $M^\mathrm{nd}(t)$ and $M^\mathrm{d}(t)$, given by
Eqs.~(\ref{S32}) and (\ref{S36}) respectively. We first argue that, unlike in
the case of a global Hamiltonian perturbation, {\it the non-diagonal
  contribution will typically dominate over the diagonal term}. The most
favorable regime to observe the diagonal term would be that of perturbations
satisfying $\lambda < \kappa \gamma$. Therefore, the necessary requirement for
resolving the diagonal term is $\lambda < 4\gamma$. Since $\lambda - \gamma =
h_{\mathrm{KS}} > 0$ (with $h_{\mathrm{KS}}$ the Kolmogorov-Sinai entropy of
the chaotic repeller \cite{kantz}) we see that $M^\mathrm{d}(t)$ can possibly
prevail over $M^\mathrm{nd}(t)$ only if $h_{\mathrm{KS}} < 3\lambda/4$.
Taking into account that $h_{\mathrm{KS}} = \lambda$ for a closed system, we
see that only relatively open cavities would allow to observe
$M^\mathrm{d}(t)$.  Hence for further analysis of the LE decay we mainly focus
on the non-diagonal contribution $M^\mathrm{nd}(t)$.

According to Eq.~(\ref{S32}) the LE displays different decay regimes depending
on the strength of the perturbation. Thus, for weak perturbations
characterized by $\chi \lesssim 1$ one can expand the phase factor
$e^{iu/\lambdabar}$ in the Taylor series to the second order in $u/\lambdabar$
to obtain $\Re \langle e^{iu/\lambdabar} \rangle \approx 1 - \chi^2/2$ and,
therefore,
\begin{equation}
  M^\mathrm{nd}(t) \approx e^{-\chi^2 \gamma t} \;\;\;\;\; $for$
  \;\;\;\;\; \chi \lesssim 1 \;\;\; $(FGR)$ \, .
  \label{S39-FGR}
\end{equation}
The rate of the exponential decay given by Eq.~(\ref{S39-FGR}) depends on the
perturbation strength $\chi$, in analogy to the Fermi-golden-rule regime found
for global perturbations, but is dressed with $\gamma$ that provides a measure
of the fraction of phase-space affected by the boundary deformation. On the
other hand, in the limit of strong perturbations, $\chi \gg 1$, the LE decay
rate is independent of the perturbation strength $\chi$ and is entirely
determined by the extent of the deformation quantified by $\gamma$. Indeed,
$\langle e^{iu/\lambdabar} \rangle \rightarrow 0$ as $\chi \rightarrow \infty$
leading to
\begin{equation}
  M^\mathrm{nd}(t) \approx e^{-2\gamma t} \;\;\;\;\; $for$
  \;\;\;\;\; \chi \gg 1 \;\;\; $(escape-rate)$ \, .
  \label{S39-escape}
\end{equation}
This is the escape-rate dominated decay regime previously reported in
Ref.~\cite{gouss}. We finally emphasize that Eq.~(\ref{S32}), and therefore
Eqs.~(\ref{S39-FGR}) and (\ref{S39-escape}), hold only if the conditions
(\ref{condition-1}) and (\ref{condition-2}) are satisfied, i.e.
\begin{equation}
  \lambdabar \ll \sigma \ll \sqrt{\lambdabar\lL} \; ,
  \label{condition-combined}
\end{equation}
where $\lL$ is the Lyapunov length.

\begin{figure}[h]
\centerline{\epsfig{figure=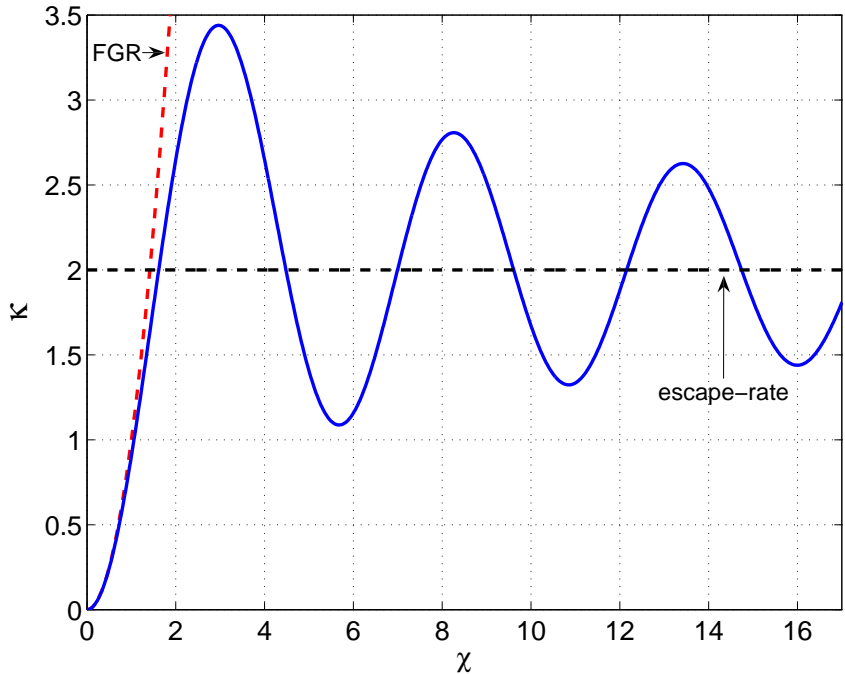,width=4.5in}}
\caption{Decay rate (in units of $\gamma$) of the Loschmidt echo in a chaotic
  billiard (Eq.~(\ref{kappa})) as a function of the perturbation strength
  $\chi$. We have chosen a piston-like boundary deformation (see
  \ref{app_piston}) for which the LE decay follows Eq.~(\ref{S32-piston}) and
  $\chi = (8/3)^{1/2}\,h/\lambdabar$. The Fermi-golden-rule decay regime
  (Eq.~(\ref{S39-FGR})) is recovered for $\chi \lesssim 1$ (red dashed line).
  As $\chi \rightarrow \infty$ the decay rate asymptotically approaches the
  value 2 (black dashed line) representing the escape-rate regime
  (Eq.~(\ref{S39-escape})).}
\label{fig-3}
\end{figure}

As a particularly interesting feature, the transition between the FGR and
escape-rate regime is non-monotonic, i.e.  the LE decay rate $\kappa \gamma$,
in general, oscillates as a function of the perturbation strength $\chi$ while
approaching the asymptotic value $2\gamma$. Figure~\ref{fig-3} illustrates
these distinct oscillations for the case of a chaotic billiard with the
Hamiltonian perturbation generated by a piston-like boundary deformation, see
\ref{app_piston}. In other words, for a fixed time $t$ and starting in a
minimum, the LE can increase by orders of magnitude upon varying the
perturbation strength $\chi$.  The strength of the oscillations depends on the
geometry of the boundary perturbation. The oscillations are particularly
pronounced for the piston-type geometry where they only very slowly merge into
the escape-rate limit.

\section{Numerics versus semiclassical predictions}
\label{section_numerics}

\subsection{Finite-size corrections to the semiclassical limit}

In deriving our approximate expression for the non-diagonal contribution to
the LE (Eq.~(\ref{S32})) we made the following two important assumptions.
Firstly, we restricted our discussion only to those initial wave packets whose
dispersion $\sigma$ is much larger than the de Broglie wavelength
$\lambdabar$, see Eq.~(\ref{condition-2}). In numerical simulations, however,
one can only address finite $\sigma/\lambdabar$ ratios, so that the theory
must be improved accordingly to be capable of accounting for the results of
the simulations. Secondly, in Sec.~\ref{section_theory} we used the simple
expression, given by Eq.~(\ref{eq-rho}), for the probabilities $\rho_n(l)$ for
a classical trajectory of length $l$ to visit the deformation region $n$
times. While this expression well approximates $\rho_n(l)$ for $l \gg \lD$ an
improved formula is needed to correctly treat the numerically assessable range
of lengths, $l \sim \lD$. In this subsection we address these two important
issues and present an expression for $M^{\mathrm{nd}}(t)$ appropriate for a
quantitative comparison with the results of the numerical simulations.

The non-diagonal contribution to the LE, as given by Eq.~(\ref{S19}), is given
by the square of the absolute value of the sum of overlaps $\bar{O}_n(t)$
(with $n=0,1,2,\ldots$) defined according to Eqs.~(\ref{eq-O_0-bar}) and
(\ref{eq-O_1-bar}). Now instead of using the asymptotic form of the Bessel
function $I_0$ (like in Sec.~\ref{section_wfoflp}) we keep the analysis
general by writing $\bar{O}_n(t)$ as an integral over the dimensionless
momentum variable $z=p/p_0$:
\begin{equation}
  \bar{O}_0(t) = \frac{2\sigma^2}{\lambdabar^2} \int_0^\infty dz \, z \,
  \rho_0(zp_0t/m) \, \exp\left[-\frac{\sigma^2}{\lambdabar^2} (z^2+1)\right]
  I_0\left(\frac{2\sigma^2}{\lambdabar^2}z\right) \, ,
  \label{N11-0}
\end{equation}
and
\begin{eqnarray}
  \bar{O}_n(t) = & \frac{2\sigma^2}{\lambdabar^2} \int_0^\infty dz \, z \,
  \left\langle e^{izu/\lambdabar} \right\rangle^n
  \label{N11-n} \\
  & \times \rho_n(zp_0t/m) \, \exp\left[-\frac{\sigma^2}{\lambdabar^2} (z^2+1)\right]
  I_0\left(\frac{2\sigma^2}{\lambdabar^2}z\right) \, , \;\;\;
  n\ge 1 \, .
  \nonumber
\end{eqnarray}

We now address the probability distributions $\rho_n(l)$. The central building
block of our analysis here is the probability $g(l) dl$ for a classical
trajectory with the length between $(L+l)$ and $(L+l+dl)$ to end on the
boundary deformation region $\mathcal{B}_1$ subject to the condition that the
previous encounter with $\mathcal{B}_1$ took place at length $L$. Assuming
ergodicity for the billiard system under consideration we approximate $g$ by
the Heaviside step function $\theta$ as
\begin{equation}
  g(l) \approx \frac{1}{\lD} \, \theta(l-l_0) \, ,
  \label{N12-0}
\end{equation}
where the length $l_0$ is the (average) minimal length that a trajectory
starting on $\mathcal{B}_1$ must have to return to $\mathcal{B}_1$. In view of
Eq.~(\ref{N12-0}) one readily obtains the following approximation for the
survival probability:
\begin{equation}
  \rho_0(l) \approx \theta(l_0-l) + \theta(l-l_0) \, e^{-(l-l_0)/\lD} \, .
  \label{N12}
\end{equation}
Then, the visit probabilities for $n\ge 1$ are calculated as
\begin{eqnarray}
  \rho_n(l)  = & \int_0^l dl_n \int_0^{l_n} dl_{n-1} \ldots \int_0^{l_2} dl_1 \,
  \rho_0(l-l_n) 
  \label{N13} \\
  & \times g(l_n-l_{n-1}) \, \rho_0(l_n-l_{n-1}) \ldots g(l_2-l_1) \,
  \rho_0(l_2-l_1) \, g(l_1) \, \rho_0(l_1) \, ,
  \nonumber
\end{eqnarray}
and can be shown to satisfy the following recursion relation for $n \ge 1$,
\begin{eqnarray}
  \rho_n(l) & = \theta(l-nl_0) \, \theta((n+1)l_0-l) \left( 1-\sum_{k=0}^{n-1}
    \rho_k(l) \right)
  \label{N14} \\
  & + \theta(l-(n+1)l_0) \left[ \left( 1-\sum_{k=0}^{n-1}
    \rho_k((n+1)l_0) \right) \right.
  \nonumber \\
  & \left. + \sum_{k=0}^{n-1} \frac{\rho_k((k+1)l_0)}{(n-k)!}
  \left( \frac{l-(n+1)l_0}{\lD} \right)^{n-k} \right] \exp\left(
  -\frac{l-(n+1)l_0}{\lD} \right) \, .
  \nonumber
\end{eqnarray}
Note that Eq.~(\ref{N14}) together with (\ref{N12}) simplifies to
Eq.~(\ref{eq-rho}) if one puts $l_0 = 0$. However, for trajectories of lengths
comparable to $\lD$ the minimal return length $l_0$ can not be neglected and
Eqs.~(\ref{N12}) and (\ref{N14}) must be used in Eqs.~(\ref{N11-0}) and
(\ref{N11-n}) to yield the LE.

\subsection{Numerical simulations}

\begin{figure}[h]
\centerline{\epsfig{figure=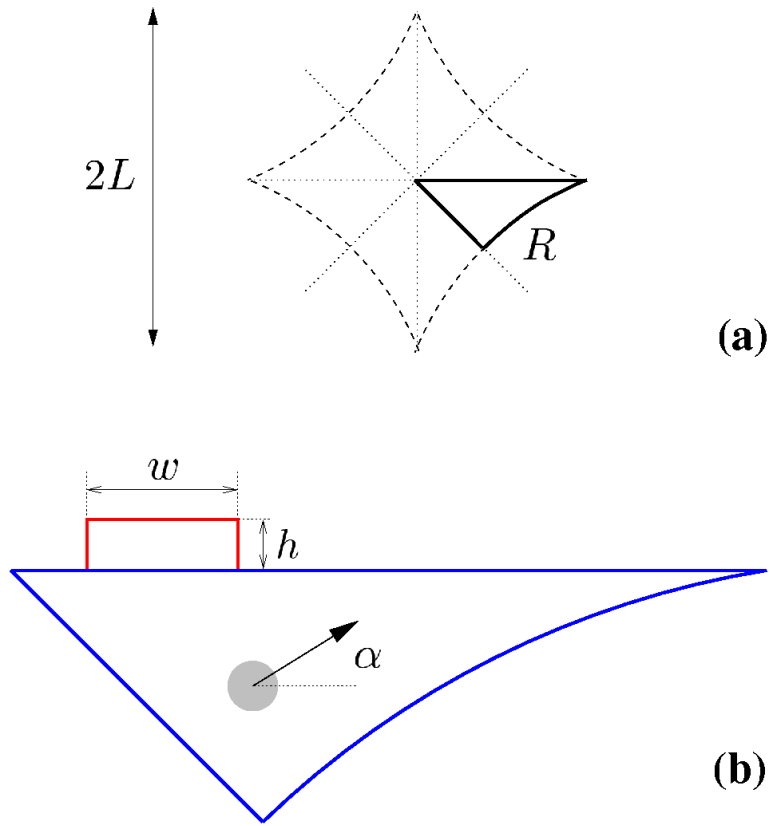,width=4.0in}}
\caption{Desymmetrized diamond billiard: (a) the fundamental domain of the
  four-disk billiard, (b) the initial wave packet (with momentum enclosing an
  angle $\alpha$ with the horizontal) in the case of a local piston-like
  boundary deformation (defined by a width $w$ and displacement $h$).}
\label{fig-4p1}
\end{figure}

In order to support our semiclassical predictions we present in this section
numerical quantum mechanical calculations for a local perturbation. We use the
Trotter-Suzuki algorithm \cite{raedt,cucch-thesis} to simulate the dynamics of
a Gaussian wave packet inside a desymmetrized diamond billiard (DDB). The DDB
is defined as a fundamental domain of the area confined by four intersecting
disks of radius $R$ centered at the vertices of a square. We denote the length
of the largest straight segment of the DDB by $L$ (see Fig.~\ref{fig-4p1}). As
proved in Ref.~\cite{szasz}, the DDB is fully chaotic and thus has been
previously considered for studying aspects of quantum chaos \cite{mull}.

Our semiclassical analysis is valid for an arbitrarily shaped local
perturbation acting on a region ${\cal B}_1$ (of width $w$) of the boundary.
A perturbation with the shape of a circular segment was used in
Ref.~\cite{gouss}. In our present numerical simulations we chose a piston-like
perturbation (Fig.~\ref{fig-4p1}b), for which analytical results can be
readily obtained (see Eqs.~(\ref{eq:piston}) and (\ref{S32-piston}), and
\ref{app_piston}).

\begin{figure}[h]
\centerline{\epsfig{figure=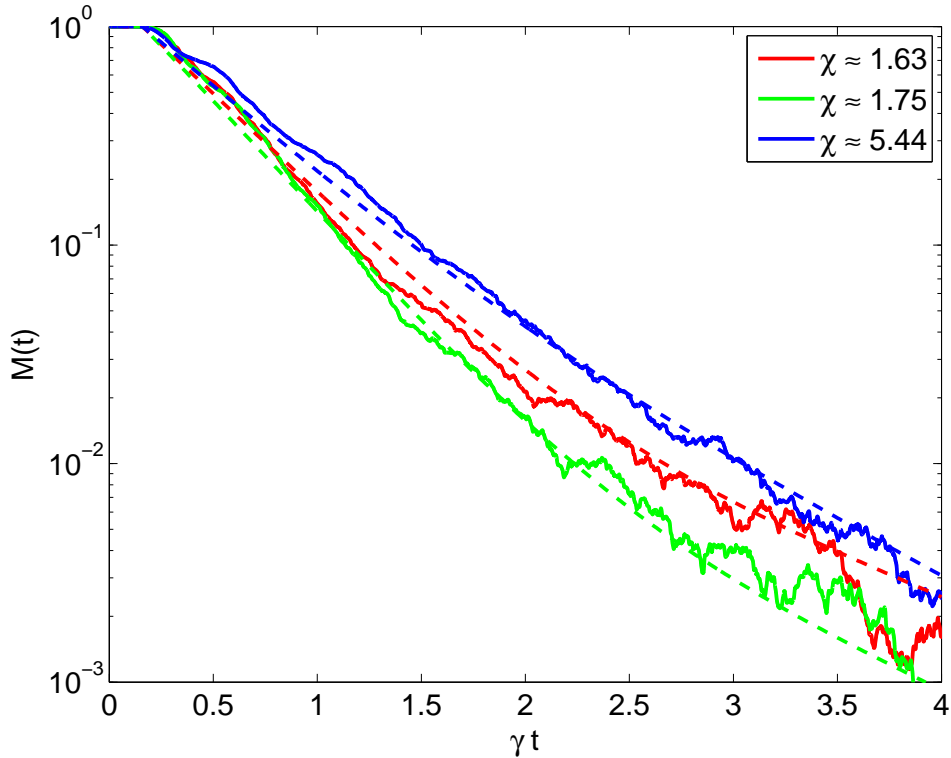,width=5.0in}}
\caption{The Loschmidt echo decay in the DDB for tree different values of the
  piston-like deformation strength (see Fig.~\ref{fig-4p1}): (i) $\chi \approx
  1.63$ ($h=3$ and $\lambdabar=3$), (ii) $\chi \approx 1.75$ ($h=3$ and
  $\lambdabar=2.8$), and (iii) $\chi \approx 5.44$ ($h=10$ and
  $\lambdabar=3$). The solid-line curves are obtained as the result of
  numerical simulations, whereas the dashed-line curves show the semiclassical
  predictions (see text). The initial wave packet dispersion is $\sigma = 8$,
  and the other system parameters are $L = 1000$, $R = 1311$ and $w = 120$.
  The time is given in units of the dwell time $1/\gamma$, with $\gamma$
  defined in Eq.~(\ref{S41}).}
\label{fig-5}
\end{figure}

In Fig.~\ref{fig-5} we present the LE decay for a DDB with $L = 1000$, $R =
1311$ (in units of the lattice spacing of the underlying tight-binding model)
and a piston-like perturbation. For the present geometry the dwell length is
$\lD = (P/w) \lF \approx 18.7 \, \lF$, with $\lF$ being the mean free flight
path. The initial wave packet has the dispersion $\sigma = 8$; its momentum
direction is chosen to be parallel to the longest straight segment of the DDB
($\alpha=0$ in Fig.~\ref{fig-4p1}.b), but we have verified that the LE decay
rate is independent of $\alpha$ \cite{footnote-3}. The three numerical
(solid-line) curves in Fig.~\ref{fig-5} correspond to the following values of
the deformation strength: (i) $\chi \approx 1.63$ ($h=3$ and $\lambdabar=3$),
(ii) $\chi \approx 1.75$ ($h=3$ and $\lambdabar=2.8$), and (iii) $\chi \approx
5.44$ ($h=10$ and $\lambdabar=3$). These numerically obtained LE curves decay
almost exponentially for times up to $\gamma t \approx 4$ before turning over
to a regime with strong irregular fluctuations around a saturation value
\cite{peres}. The three curves shown illustrate the non-monotonous dependence
of the decay rate with the perturbation strength.

The corresponding semiclassical LE decay curves -- the three dashed-line
curves in Fig.~\ref{fig-5} -- were obtained by doing the integrals in the
right-hand side of Eqs.~(\ref{N11-0}) and (\ref{N11-n}) numerically, with
$\rho_n(l)$ probability distributions determined in accordance with
Eqs.~(\ref{N12}) and (\ref{N14}); the minimal return length was taken to be
$l_0 = 0.16 \, \lD$. The infinite sum in the right-hand side of
Eq.~(\ref{S19}) was terminated at $n=8$.

The good agreement between the semiclassical predictions and the results of
the full quantum-mechanical computation is evident in Fig.~\ref{fig-5}. The
fact that the obtained LE decay deviates from the purely exponential one is
entirely due to the finiteness of the $\sigma/\lambda$ ratio: the analytical
results of Sec.~\ref{section_theory} are recovered in the limit given by
Eq.~(\ref{condition-combined}), which proves challenging in numerical
simulations \cite{footnote-4}. As we discuss in the next section, however,
this limit is naturally achievable in laboratory experiments with ultra-cold
atom-optics billiards, so that the latter provide a viable model system for
studying the LE from local Hamiltonian perturbations.

\section{ Experimental realizations of Loschmidt echo with a local
perturbation}
\label{section_experiment}

Experiments on the LE are of foremost importance since they render crucial
information about quantum dynamics of physical systems and their decoherence
mechanisms \cite{experiments}. While the examples discussed in the
introduction show that the agreement between the semiclassical theory of the
LE and numerical simulations is quite successful, the situation is less
satisfactory concerning experiments.

LE experiments were first performed on nuclear spins of organic molecules
using NMR techniques \cite{cit-MolPhysics,cit-PhysicaA}.  The decay of the
polarization was found to be quite insensitive to the coupling to external
degrees of freedom or the precision of the reversal.  The Gaussian decay of
the experimentally measured LE is at odds with the one-body semiclassical
theory, and many-body aspects of the problem have been pointed to be at the
origin of such a behavior \cite{pros-NEW,Zurek_Cuc,jacq-3}.

In Ref.~\cite{gouss} a principle experimental scheme for measuring the LE for
local boundary perturbations was proposed based on a ballistic electron cavity
with a small ferromagnet attached acting as the local perturbation. Such a
setting provides a link between spin relaxation in a mesoscopic cavity and LE
decay.  Here we discuss two further experimental settings which appear
suitable for a measurement of the echo decay: microwave and cold atom
cavities.

Microwave experiments allow the independent measurement of individual
scattering matrix elements for the unperturbed and perturbed systems
\cite{Stoeckmann}. The cross-correlation of these matrix elements can then be
calculated, and going into the time domain, the scattering fidelity is
obtained. The latter is a good representation of the usual average fidelity
amplitude when appropriate ensemble and/or energy averages are taken.
Correspondingly, the LE can also be constructed from measurements of
scattering matrix elements. The observation of the Lyapunov decay regime for a
global perturbation in the microwave cavity can then be envisioned. However,
reaching the required long time scales remains as an experimental challenge.
On the other hand, the corresponding (escape-rate) decay regime for local
perturbations might be easier to reach experimentally. Moreover, microwave
billiards appear to be rather suited to investigate local boundary
perturbations, since the piston-type deformation presented in our work can be
directly realized in a microwave billiard setup.  The width of the piston
determines the exponent of the LE time decay.  Furthermore, by moving the
piston the perturbation strength $\chi$ can be directly controlled and tuned.
Hence, by devising sufficiently large microwave cavities to approach the
semiclassical limit it seems promising to experimentally reach both, the
escape-rate regime for large $\chi$ and the non-monotonic dependence of the
decay rate of the LE on the perturbation strength (see Fig.~\ref{fig-3}).

Studying quantum chaos in the laboratory by recreating a delta-kicked harmonic
oscillator in an ion trap was proposed a decade ago by Gardiner and
collaborators \cite{Gardiner}, and a number of fruitful approaches have since
then been developed and successfully realized using ultra-cold atoms confined
to optical billiards \cite{fried}.  For instance, the decay of quantum
correlations has been measured by echo spectroscopy on ultra-cold atoms using
the detuning of the trapping laser as a perturbation \cite{anders}.  Below we
focus on the time evolution of clouds of ultra-cold atoms in optical
billiards, and show that they provide a viable system for experimental
investigation of different perturbations and various regimes of the LE decay.
The perturbations can be global, such as in the cases previously studied, but
also local. Since the large-scale separation of system parameters, given by
the condition (\ref{condition-combined}), is attainable in these experimental
systems we expect that a direct support for of the theoretical predictions of
Sec.~\ref{section_theory} can be obtained.

\begin{figure}[h]
\centerline{\epsfig{figure=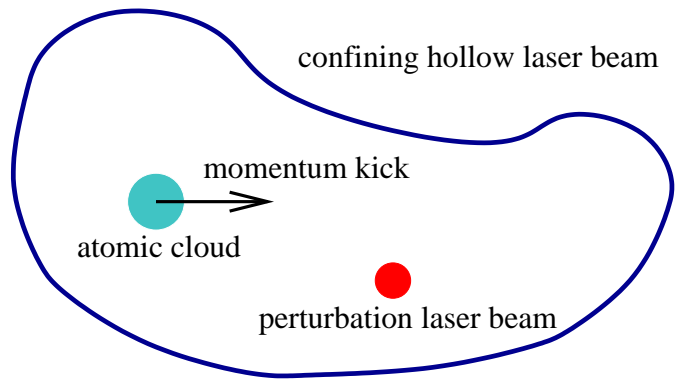,width=3.5in}}
\caption{Atom-optics billiard: a sketch of the focal plane of the hollow laser
  beam.}
\label{fig-6}
\end{figure}

In a typical microwave echo (or Ramsey) spectroscopy experiment \cite{anders}
a cloud of ultra-cold Rb atoms is loaded into an off-resonance optical trap.
For the purpose of our study the role of the trap can be played by a hollow
laser beam with the cross section corresponding to the geometry of a chaotic
billiard of interest. The fabrication of such hollow laser beams, as well as
the manipulation of atoms inside them, can now be performed with a high level
of precision \cite{arak}. The atomic cloud, after being positioned inside the
hollow beam in its focal plane and accelerated (or ``kicked'') as a whole to a
nonzero average momentum, is let to evolve freely in an effectively
two-dimensional billiard, see Fig.~\ref{fig-6}.

The Rb atoms used in echo experiments \cite{anders} are initially prepared in
a quantum state $\Psi_0$ equal to a direct product of an internal atomic state
$|s\rangle$ and a spatial state described by a wave function $\phi_0(\vr)$,
i.e. $\Psi_0= |s\rangle \otimes \phi_0(\vr)$. The internal state evolves in a
coherent superposition of the two hyperfine sub-states, denoted by $\sd$ and
$\su$, of the ground state of rubidium. The $\sd$-component of the total
wave function of an atom experiences a laser field potential
$V_\downarrow(\vr)$ which is, in general, different from the potential
$V_\uparrow(\vr)$ exerted by the same laser on the $\su$-component. The
relative difference between the two optical potentials is given by the ratio
$\omega_\mathrm{HF}/\Delta_\mathrm{L}$, where $\hbar\omega_\mathrm{HF}$ is the
energy of the hyperfine splitting of the ground state, and $\Delta_\mathrm{L}$
is the laser detuning from the frequency of the transition between the ground
state and the first allowed excited state of the Rb atom. The application of a
sequence of $\pi/2$ microwave pulses during the time evolution of the atoms,
followed by a measurement of the populations of the $\sd$- and
$\su$-sub-states at the end of the evolution, allows one to determine the LE
(corresponding to the spatial wave function $\phi_0(\vr)$) due to the
difference between the potentials $V_\downarrow(\vr)$ and $V_\uparrow(\vr)$ as
a function of the evolution time.

In order to measure the LE decay due to local perturbations two different
lasers have to be used. The first laser is to produce the confining hollow
beam with the cross section of a desired (billiard)
geometry, and has to be tuned as to exert approximately the same potential
$V_\mathrm{bill}(\vr)$ on the both $\sd$- and $\su$-sub-states. The beam of
the second laser plays a role of the local Hamiltonian perturbation. It has to
be placed inside (and aligned with) the hollow beam of the first confining
laser, and its width should be much smaller than the linear scale of the
billiard (see Fig.~\ref{fig-6}). The frequency of this second laser (and
perhaps its position inside the billiard) determines the
perturbation strength $\chi$. Altering this frequency changes the difference
between the potentials $\delta V_\downarrow(\vr)$ and $\delta
V_\uparrow(\vr)$, produced by the second laser and acting differently on the $\sd$- and
$\su$-substates, respectively. Thus, an echo spectroscopy experiment performed
in such a system would measure the LE decay due to the difference of the atomic
potentials $V_\downarrow(\vr)=V_\mathrm{bill}(\vr)+\delta V_\downarrow(\vr)$
and $V_\uparrow(\vr)=V_\mathrm{bill}(\vr)+\delta V_\uparrow(\vr)$; this
difference is localized in an area much smaller than that of the billiard.

To date one is typically able to experimentally prepare and manipulate clouds
of Rb atoms as cold as 1 $\mu$K. This temperature corresponds to the thermal
speed of about 1.3 cm/s. At the same time, by first placing the atoms inside a
far-off-resonance dipole trap then moving the trap and finally switching it
off one can accelerate the atomic cloud as a whole up to 10 cm/s. Such a
momentum kick can nowadays be easily realized in a laboratory, and does not
significantly increase the temperature of the atoms.  As a result one obtains
a cloud of atoms moving as a whole with an average momentum that corresponds
to a rescaled de Broglie wave length $\lambdabar \sim$ 10 $n$m. The dispersion
of the atomic cloud can be shrunk to $\sigma \approx$ 1 $\mu$m. Under this
conditions the number of Rb atoms composing the cloud can reach $10^5$ that is
well sufficient for the Ramsey-spectroscopy-type measurements.  The hollow
laser beam, producing the billiard confinement, can reach $L \approx$ 1 cm in
linear size.  Assuming the Lyapunov scale $\lL$ of the billiard to be of the
same order of magnitude as $L$ one arrives at the following estimate for the
scale separation of the system parameters:
\begin{equation}
  \lambdabar : \sigma : \sqrt{\lambdabar\lL} \sim 1 : 10^2 : 10^3,
  \label{C11}
\end{equation}
which well satisfies the restriction given by Eq.~(\ref{condition-combined}).
Under the conditions specified above one can control the atoms for up to 5 s
before the cloud gets significantly elongated in the axial direction of the
hollow laser beam. Such a time allows a single atom to experience about $50$
bounces with the billiard boundary, which according to our theory is
sufficient for observing the LE decay regimes predicted in this work.

The above considerations show that atom-optics billiards constitute promising
candidates for experimentally investigating different regimes of the LE decay
due to local perturbations of the Hamiltonian, and in particular put in
evidence the escape-rate regime and the predicted oscillations of the LE as a
function of the perturbation strength.

\section{Conclusions}
\label{section_conclusion}

In this work we have studied the time decay of the Loschmidt echo in quantum
systems that are chaotic in the classical limit, due to Hamiltonian
perturbations localized in coordinate space. We have provided the
corresponding semiclassical theory of the LE for coherent initial states
evolving in two-dimensional chaotic billiards.

In addition to the FGR decay regime, which is well-known for the case of
global Hamiltonian perturbations and is recovered in our theory for weak
($\chi \lesssim 1$) local perturbations, our analysis predicts a novel decay
regime for strong ($\chi \gg 1$) perturbations that stems entirely from the
local nature of the Hamiltonian perturbation, i.e. the {\it escape-rate}
regime, and quantitatively describes the transition between the FGR and
escape-rate regimes as the perturbation strength is varied. In the escape-rate
regime the LE decays exponentially in time with a rate equal to twice the
escape rate from an open billiard with the ``hole'' at the place of the
perturbation. Hence the LE allows to mimic the decay behavior of a system
without opening it. In this regime the LE decay rate is independent of the
deformation strength $\chi$. The transition between the FGR regime and the
escape-rate regime turns out to be non-monotonic: the rate of the exponential
time-decay of the LE oscillates as a function of the perturbation strength.

We would like to point out that recently there has been another study
\cite{hoehmann} of the LE decay due to local perturbations. It addresses a
particle moving in a two-dimensional array of point-like scatterers, and the
perturbation of the Hamiltonian is achieved by slightly displacing one of the
scatterers. The theoretical and experimental analysis of the system reveals a
polynomial decay of the LE, namely $M(t) \sim t^{-2}$, for long times, which
cannot be compared with our findings: that work is in the perturbative regime,
where the eigenstates are not significatively modified by the perturbation.

We have also performed an extensive numerical study of the LE decay to support
our semiclassical theory. To this end we have simulated the time evolution of
initially coherent states in the DDB. The role of the local Hamiltonian
perturbation was played by a piston-like deformation of the billiard boundary.
The results of our numerical simulations exhibit strong quantitative agreement
with the predictions of our theory extended to cope with initial states given
by Gaussian wave packets of a dispersion comparable with the de Broglie
wavelength.

While the scale separation given by Eq.~(\ref{condition-combined}) is rather
challenging to be satisfied numerically it can be naturally achieved in
laboratory experiments with ultra-cold atoms confined to optical billiards. In
this work we have proposed a laboratory set-up allowing one to investigate the
LE decay from local Hamiltonian perturbations for a wide range of perturbation
strengths, and to observe the predicted decay regimes. We believe that the
study of the LE decay due to local perturbations provides an example of
physical problems for which capabilities of laboratory experiments go beyond
those of numerical simulations. Such experiments may also reveal
weak-localization-type quantum corrections to the LE decay which are expected
from an analysis \cite{Waltner} of loop contributions \cite{Sieber} beyond the
semiclassical diagonal approximation.

\ack The authors would like to thank Inanc Adagideli, Arnd B\"acker, Philippe
Jacquod, Thomas Seligman and Steven Tomsovic for helpful conversations.
Special acknowledgments go to Ilya Arakelyan for providing the authors with
helpful information about state-of-the-art atom-optics experiments. This work
has been supported by the Alexander von Humboldt Foundation (AG), the
Vielberth foundation (RJ) and the Deutsche Forschungsgemeinschaft within FG
760 (RJ, KR) and GRK 638 (DW, KR).

\appendix

\section{Linearization of the action integral}
\label{app_action}

Here we present the details of the expansion of the action integral (or the
Hamilton principal function) around the central trajectory of the wave packet.
This approximation is an important aspect of the semiclassical description of
the LE (see Ref.~\cite{cucch-1} and Sec.~\ref{section_theory} of the present
work), and its limits require careful consideration. Thus, we devote this
appendix to the validity condition of the linear expansion. Higher order
expansions have been recently considered in the literature
\cite{wang,vanic,wang2}.

As shown in Fig.~\ref{fig-1}, we consider a wave packet centered at $\vr_0$
and localized at a small circular region of radius $\sigma$. The action
integral $S_{\hat{s}}(\vr,\vr',t)$ along a trajectory $\hat{s}$ starting at
a point $\vr'$ within this circular region at time
$0$ and leading to $\vr$ in a time $t$ can be expanded as
\begin{eqnarray}
  S_{\hat{s}}(\vr,\vr',t) = S_{s}(\vr,\vr_0,t) &+ (\vr'-\vr_0) \cdot
  \left[ \frac{\partial S_{\hat{s}}(\vr,\vr',t)}{\partial\vr'}
  \right]_{\vr'=\vr_0}
  \label{A1.1}\\
  &+ \frac{1}{2} (\vr'-\vr_0) \cdot \left[
    \frac{\partial^2 S_{\hat{s}}(\vr,\vr',t)}{\partial\vr'^2}
  \right]_{\vr'=\vr_0} (\vr'-\vr_0) + \ldots \; .
  \nonumber
\end{eqnarray}
Here we assume that the trajectory
$\hat{s}(\vr,\vr',t)$ converges to the central trajectory
$s(\vr,\vr_0,t)$ as $\vr'$ approaches $\vr_0$. The dot denotes the
scalar (as opposed to matrix) multiplication. Using the identity
$\partial S_{\hat{s}} / \partial\vr' = -\vp_{\hat{s}}$, where
$\vp_{\hat{s}}$ denotes the initial momentum on the trajectory
$\hat{s}(\vr,\vr',t)$, we rewrite Eq.~(\ref{A1.1}) as
\begin{eqnarray}
  S_{\hat{s}}(\vr,\vr',t) = S_{s}(\vr,\vr_0,t) &- \vp_s \cdot
  (\vr'-\vr_0)
  \label{A1.2}\\
  &- \frac{1}{2} (\vr'-\vr_0) \cdot \left[
    \frac{\partial\vp_{\hat{s}}}{\partial\vr'} \right]_{\vr'=\vr_0}
  (\vr'-\vr_0) + \ldots \; .
  \nonumber
\end{eqnarray}
Note that in our notation $\vp_{\hat{s}} \rightarrow \vp_s$ as $\vr'
\rightarrow \vr_0$, see Fig.~\ref{fig-1}. In order to
truncate the expansion (\ref{A1.2}) at the  term linear in
$\vr'-\vr_0$, and therefore recover Eq.~(\ref{S4}), the 
condition 
\begin{equation}
  \left| (\vr'-\vr_0) \cdot \left[
      \frac{\partial\vp_{\hat{s}}}{\partial\vr'} \right]_{\vr'=\vr_0}
    (\vr'-\vr_0) \right| \ll \hbar
  \label{A1.3}
\end{equation}
must be satisfied for all points $\vr'$ such that $|\vr'-\vr_0| \lesssim \sigma$.

To analyze Eq.~(\ref{A1.3}) we introduce  a system of relative coordinates
moving along the central trajectory $s(\vr,\vr_0,t)$. Thus, for any
time $\tau \in [0,t]$ the distance between the phase space points
$({\bf q}'_\tau,\vp'_\tau)$ and $({\bf q}_\tau,\vp_\tau)$, belonging
to the trajectory $\hat{s}(\vr,\vr',t)$ and $s(\vr,\vr_0,t)$
respectively, is given by ${\bf q}'_\tau - {\bf q}_\tau =
(q_\tau^\parallel, q_\tau^\perp)$ and $\vp'_\tau - \vp_\tau =
(p_\tau^\parallel,p_\tau^\perp)$, where the superscripts 
$\parallel$ and $\perp$ refer to the vector components parallel and
perpendicular to $\vp_\tau$. (Note that in the current notation ${\bf
q}_0 \equiv \vr_0$, ${\bf q}'_0 \equiv \vr'$, ${\bf q}_t = {\bf q}'_t
\equiv \vr$, $\vp_0 \equiv \vp_s$ and $\vp'_0 \equiv \vp_{\hat{s}}$.)
Then
\begin{equation}
  \left[ \frac{\partial\vp_{\hat{s}}}{\partial\vr'}
  \right]_{\vr'=\vr_0} = \left(
    \begin{array}{cc}
      \partial p_0^\parallel / \partial q_0^\parallel & \partial p_0^\parallel / \partial q_0^\perp\\
      \partial p_0^\perp / \partial q_0^\parallel & \partial p_0^\perp / \partial q_0^\perp
    \end{array} \right)_{(q_0^\parallel, \, q_0^\perp) = {\bf 0}}.
  \label{A1.4}
\end{equation}
For a billiard the off-diagonal partial derivatives can be
neglected compared to the diagonal ones, so that condition~(\ref{A1.3}) can be
replaced by
\begin{equation}
  \sigma^2 \left| \frac{\partial p_0^\parallel}{\partial
  q_0^\parallel} + \frac{\partial p_0^\perp}{\partial
  q_0^\perp}\right|_{(q_0^\parallel, \, q_0^\perp) = {\bf 0}} \ll 
  \hbar \; .
  \label{A1.5}
\end{equation}
The first of the two derivatives in Eq.~(\ref{A1.5}) is 
$\partial p_0^\parallel / \partial q_0^\parallel = -m/t$
for a particle of mass $m$ in a billiard.  To evaluate the second derivative 
we first linearize the trajectory $\hat{s}(\vr,\vr',t)$  around $s(\vr,\vr_0,t)$, 
so that $q_\tau^\perp \approx q_\tau^\perp(q_0^\perp, p_0^\perp, \tau)$.
Therefore,
\begin{equation}
  0 \equiv dq_t^\perp = \left( \frac{\partial q_t^\perp}{\partial
      q_0^\perp} \right)_{p_0^\perp} dq_0^\perp + \left( \frac{\partial
      q_t^\perp}{\partial p_0^\perp} \right)_{q_0^\perp} dp_0^\perp \; ,
  \nonumber
\end{equation}
which leads to
\begin{equation}
  \left(\frac{\partial p_0^\perp}{\partial
      q_0^\perp}\right)_{q_t^\perp} = - \frac{\left(\partial
      q_t^\perp/\partial q_0^\perp\right)_{p_0^\perp}}{\left(\partial
      q_t^\perp/\partial p_0^\perp\right)_{q_0^\perp}} \; .
  \label{A1.6}
\end{equation}
The right hand side of Eq.~(\ref{A1.6}) is given by the ratio of the two
monodromy matrix elements. To facilitate our analytical presentation, we use
here the monodromy matrix of the dynamics on Riemann surfaces of constant
negative curvature \cite{footnote-5},
\begin{equation}
  \left(
    \begin{array}{cc}
      \partial q_t^\perp / \partial q_0^\perp & \partial q_t^\perp /
      \partial p_0^\perp \\ \partial p_t^\perp / \partial q_0^\perp &
      \partial p_t^\perp / \partial p_0^\perp
    \end{array} \right) = \left(
    \begin{array}{lr}
      \cosh(\lambda t) & (m\lambda)^{-1}\sinh(\lambda t) \\
      m\lambda\sinh(\lambda t) & \cosh(\lambda t)
    \end{array} \right),
  \label{A1.7}
\end{equation}
with $\lambda$ the Lyapunov exponent. For times
longer than the Lyapunov time, $t \gg 1/\lambda$, we have $\partial
p_0^\perp / \partial q_0^\perp \approx -m\lambda$, so that
Eq.~(\ref{A1.5}) can be replaced by
\begin{equation}
  \sigma^2 m\lambda \ll \hbar \, .
  \label{A1.8}
\end{equation}
In terms of the Lyapunov length $\lL = (p/m)(1/\lambda)$,
conveniently used for billiards, Eq.~(\ref{A1.8}) then reads
\begin{equation}
  \sigma \ll \sqrt{\frac{\hbar\lL}{p}} \; .
  \label{A1.9}
\end{equation}
The momentum uncertainty of a Gaussian wave packet of dispersion $\sigma$ is
$\hbar/\sigma$, so that $p \lesssim \hbar/\lambdabar + \hbar/\sigma$, with
$\lambdabar$ being the rescaled de Broglie wavelength. Therefore, condition
(\ref{A1.9}) holds for every trajectory relevant for the wave packet
propagation only if
\begin{equation}
  \sigma \ll \sqrt{\frac{\lL}{1/\lambdabar+1/\sigma}} \; .
  \label{A1.10}
\end{equation}
The action integral expansion (\ref{S4}) requires 
condition (\ref{A1.10}) to be satisfied. We
finally note that in the limit $\lambdabar \ll \sigma$, which we utilize in
Sec.~\ref{section_theory}, Eq.~(\ref{A1.10}) simplifies to $\sigma \ll
\sqrt{\lambdabar\lL}$.

\section{Piston-like boundary deformation}
\label{app_piston}

\begin{figure}[h]
\centerline{\epsfig{figure=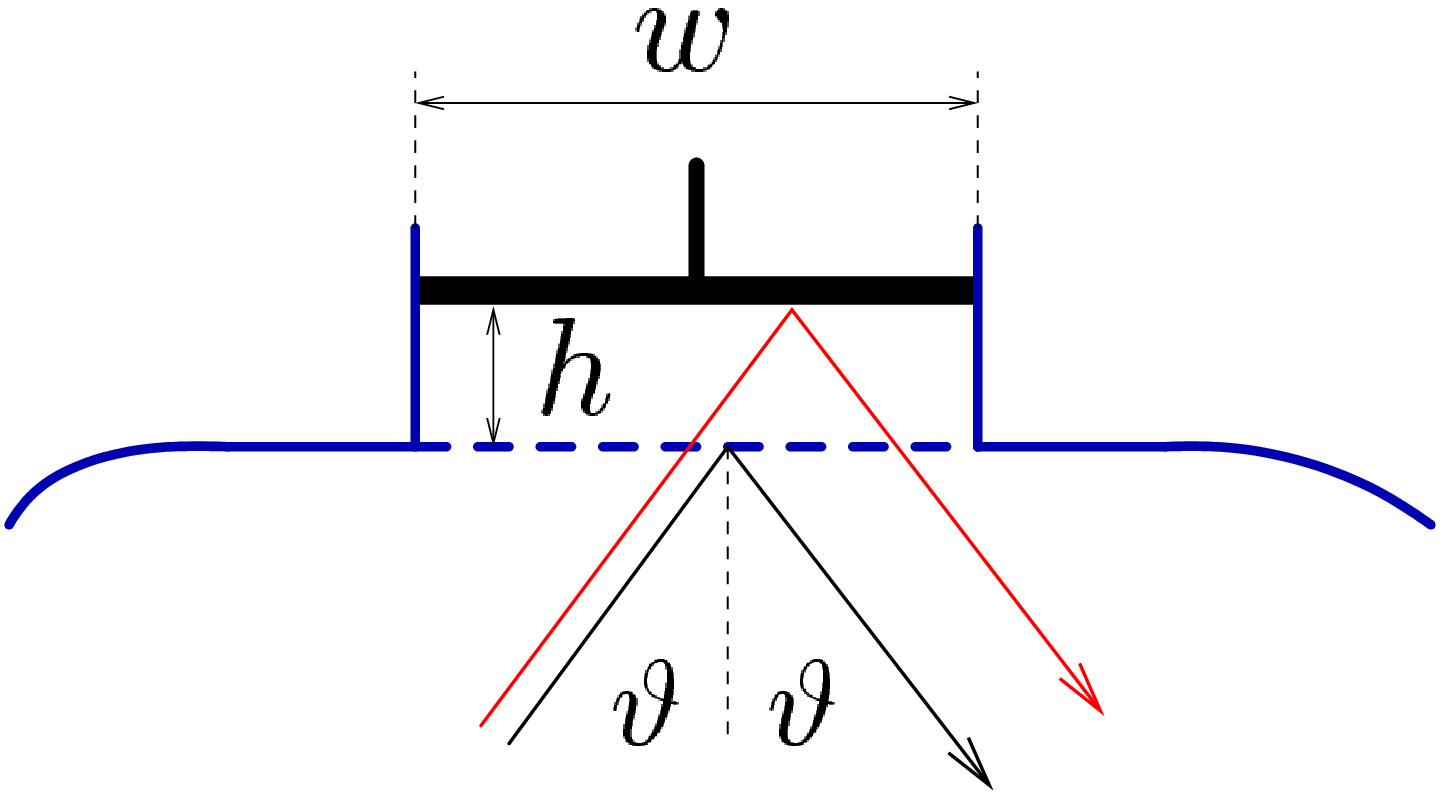,width=3.5in}}
\caption{Piston-like boundary deformation.}
\label{fig-7}
\end{figure}

In this appendix we explicitly compute as an example the length-difference function
$u(\vartheta,\xi)$ of Eq.~(\ref{S24}) for a piston-like local boundary
deformation, see Fig.~\ref{fig-7}, which is also used in our numerics.
We assume that the boundary of the unperturbed billiard
possesses a straight segment of length $w$ that gets
``lifted'' by the perturbation as if an imaginary ``piston'' was pulled out.
We denote the piston displacement by $h$.
Assuming $h$ much smaller than the free
flight path $\lF$ of the trajectory hitting the deformation, 
we treat the unperturbed and perturbed trajectories to be parallel.
Then the length difference $u(\vartheta,\xi)$
accumulated due to a single collision with the deformation-affected segment of
the boundary is given by an expression analogous to Bragg's diffraction
formula,
\begin{equation}
  u(\vartheta,\xi) \approx 2h \cos\vartheta \, .
  \label{A2.1}
\end{equation}
Here $\vartheta$ represents the collision angle as shown in Fig.~\ref{fig-7}.
The length difference $u$ can be considered independent of the
collision coordinate $\xi$ for deformations such that $h \ll w$.

Taking into account the probability distribution function of collision angles,
 Eq.~(\ref{S27}), we have $\langle\cos\vartheta\rangle = \pi/4$ and
$\langle\cos^2\vartheta\rangle = 2/3$. Consequently, the first two moments of
the deformation function read
\begin{equation}
  \langle u\rangle = \frac{\pi}{2}h \qquad \mathrm{and} \qquad
  \langle u^2\rangle = \frac{8}{3}h^2.
  \label{A2.2}
\end{equation}
Similarly, the average phase factor due to a single collision of the particle
with the piston is given by Eq.~(\ref{piston-average}).
%\begin{equation}
%  \left\langle e^{i u / \lambdabar} \right\rangle = \int_0^{\pi/2} d\vartheta \cos\vartheta
%  e^{2i h \cos\vartheta / \lambdabar} = 1-\frac{\pi}{2} \left[ {\bf
%      H}_1(2h/\lambdabar) - i J_1(2h/\lambdabar) \right] \, ,
%  \label{A2.3-new}
%\end{equation}
%where ${\bf H}_1$ is the Struve H-function of the first order, and $J_1$ is
%the first order Bessel function of the first kind.

\bigskip

%%%%%%%%%%%%%%%%%%%%%%%%%%%%%%%%%%%%%%%%%%%%%%%%%%%%%%%%%%%%%%%%%%%%%%%%%%%%%%%%


\begin{thebibliography}{99}
\bibitem{footnote-1} Note that in our definition of $\lambdabar$ there is a
  factor of $2 \pi$ with respect to the standard convention used for the de
  Broglie wavelength.
\bibitem{footnote-2} In our analysis we only consider such perturbations that
  smoothly deform the unperturbed trajectories to the perturbed ones.
\bibitem{footnote-3} We exclude initial conditions for which the wave packet
  hits the perturbation before having considerably explored the allowed phase
  space.
\bibitem{footnote-4} Scaling $\sigma \rightarrow n\sigma$ also requires $\lL
  \rightarrow n^2\lL$ to keep the ratio $\sigma/\sqrt{\lambdabar\lL}$
  unchanged. This in turn is equivalent to increasing the linear size of the
  billiard $n^2$ times.  The increase of the number of the computational grid
  points is then given by the factor $n^4$, and the number of time steps to
  propagate the wave function through the same total time multiplies by the
  factor $n^2$. This gives the overall factor $n^6$ for the code running time.
\bibitem{footnote-5} The dynamics of the motion on Riemann surfaces of
  constant negative curvature is known to be uniformly hyperbolic, with all
  trajectories possessing the same Lyapunov exponent.
\end{thebibliography}
\end{document}